\documentclass[epj]{svjour}
\usepackage{graphics}
\usepackage[colorlinks=true,linktocpage=true,linkcolor=blue,citecolor=blue,allcolors=blue]{hyperref }
\usepackage{color}
\usepackage{cite}
\usepackage{epsfig}
\usepackage{graphics}
\usepackage{amssymb}
\usepackage{verbatim}  
\usepackage{booktabs}
\usepackage{amsmath,array}

\def\lsim{\raise0.3ex\hbox{$<$\kern-0.75em\raise-1.1ex\hbox{$\sim$}}}
\def\gsim{\raise0.3ex\hbox{$>$\kern-0.75em\raise-1.1ex\hbox{$\sim$}}}
\journalname{Eur. Phys. J. C}
\begin{document}

\title{Conserved number fluctuations under global rotation in a hadron resonance gas model}

\author{Gaurav Mukherjee\inst{1,2}\thanks{phy.res.gaurav.m@gmail.com}, Dipanwita Dutta\inst{1,2}\thanks{ddutta08@hbni.ac.in}, and Dipak Kumar Mishra\inst{1}\thanks{dkmishra@barc.gov.in}
}                     
%
%
\institute{Nuclear Physics Division, Bhabha Atomic Research Centre, Mumbai - 400085, INDIA
  \and Homi Bhabha National Institute, Anushaktinagar, Mumbai - 400094, INDIA}
  
%

\abstract{
Net-baryon number, net-charge and net-strangeness fluctuations measured
in ultra-relativistic heavy-ion collisions may reveal details and insights into the 
quark-hadron transition, hadrochemical freeze-out and possibly aid 
in the search of the QCD critical point. By scanning in collision energy, 
current and upcoming heavy-ion facilities aim to explore the finite density regime 
where the critical point may lie. Effects due to rotation are also expected in 
case of peripheral collisions and we report on conserved number susceptibilities 
as calculated in the hadron resonance gas model augmented by a global angular velocity. 
Since these quantities are directly related to the experimentally measurable moments 
of the corresponding distributions our results show the possible impact of vorticity 
on the theoretical baseline and should be useful for referencing with experimental 
data and QCD-based calculations. 
} 

\authorrunning{G. Mukherjee,  D. Dutta, D. K. Mishra}
\titlerunning{Conserved number fluctuations under global rotation ...}
\maketitle

\section{\label{sec:s1} Introduction}

Properties of QCD matter under extreme conditions are being probed at Relativistic 
Heavy Ion Collider (RHIC) and Large Hadron Collider (LHC), and will be further 
studied in upcoming facilities like Nuclotron-based Ion Collider fAcility (NICA), 
JINR, Dubna and the Facility for Antiproton, Ion Research (FAIR), GSI, Darmstadt
and Heavy-Ion program at Japan Proton Accelerator Research Complex (J-PARC-HI), Japan.
In the peripheral nucleus-nucleus collisions, the created fireball may sustain rapid 
rotation for which the angular momentum is generated as a result of the initial 
non-zero impact parameter, $b$. In such heavy-ion collisions (HIC), the two colliding 
nuclei carry a total angular momentum $J \propto b\sqrt {s_{NN}}$, where $\sqrt 
{s_{NN}}$ is the nucleon-nucleon center-of-mass energy.  Though most of the total 
angular momentum is carried away by the spectators, a finite amount of the order of 
$10^4$- $10^5 \hbar$ remains in the fireball with local angular velocity in the 
range $0.01$-$0.1$ 
GeV~\cite{Becatini:2015,JiangLinLiao:2016prc,LiangWang:2016,JiangLiao:2016prl,
Deng:2016}. This scenario is experimentally supported by the STAR experiment at RHIC 
with the measurement of a non-zero value of $\Lambda$ and $\overline{\Lambda}$ 
polarization which can be translated to large values of the vorticity or angular 
velocity, $\omega \sim (9 \pm 1) \times 10^{21} s^{-1}$ $\sim 
0.05m_{\pi}$~\cite{STAR:2017}. In addition to temperature $T$ and baryon chemical 
potential $\mu_B$, vorticity or angular velocity ($\omega$) acts as an additional 
control parameter and should influence the thermodynamics and phase diagram in 
non-trivial ways~\cite{Fujimoto:2021,Fukushima:2019,Becatini:2020,Mukherjee:2023qvq}.
 
Many interesting phenomena can occur in rotating QCD matter. As examples, chiral 
vortical effect~\cite{Khar:2007,Son:2009,Khar:2011} and chiral vortical 
wave~\cite{Jiang:2015} can be induced due to fluid rotation. On the other hand,
the rotational counterpart of magnetic catalysis, i.e., the formation of the scalar 
condensate which leads to spontaneous breaking of chiral symmetry, is also being 
investigated~\cite{Ebihara:2017}. Thus, it is important to consider the effect of 
rotation while studying the properties of the medium formed in  
HICs. The first-principles study of QCD in 
rotating frames by using lattice simulations has seen recent 
interest~\cite{Yamamoto:2013, Braguta:2020}. However the notorious sign problem that 
plagues the finite density calculations returns in the case of rotating matter due 
the the effective chemical potential induced by the latter. This motivates the 
application of QCD-based models to probe dense rotating QCD matter.

The QCD phase transitions are of a crossover type at small values of $\mu_B$ which is
established by lattice QCD \cite{Aoki:2006,Aoki:2006nat,Borsanyi:2015,Peterczky:2013,Friman:2011} 
but is expected to become first order at higher values of $\mu_B$ as corroborated by 
various effective model calculations~\cite{Asakawa:1989,Barducci:1990,Berges:1999}.
The existence of a critical end point (CEP) is thus expected at the termination 
of the first-order phase transition line~\cite{Ejiri:2006,Stephanov:1999}. Locating this 
CEP is an exciting frontier in experimental as well as theoretical high energy
nuclear physics (see Ref.~\cite{Bzdak:2020} for a recent review).

The presence of CEP would lead to large correlation lengths and result 
in divergent fluctuations in various thermodynamic quantities. These may be accessed 
from the event-by-event fluctuations analyses of the quantum numbers or conserved 
charges, $viz.$  net-baryon number, net-charge and net-strangeness, obtained from 
heavy-ion collision data. This will be possible if the CEP lies close to 
the freeze-out curve and the large fluctuations survive in remnant form until 
freeze-out to a sufficient level. The moments (more accurately, cumulants) of such 
distributions have  been proposed as sensitive indicators of
a transition between hadronic and quark-gluon matter and may direct to the location
of the CEP~\cite{Karsch_2011,Gupta:2011,Gavai:2011,Karsch:2011,Garg:2017,Garg:2013}.    

The thermal properties of hot and dense QCD matter are described well by the hadron 
resonance gas (HRG) model. 
The experimental 
fits yield the freeze-out parameters and these show a consistent behavior for 
different collision energies~\cite{Cley:2006}. 
The HRG model also provides the theoretical baseline against which 
the large fluctuations and correlations in the vicinity of the critical point may be starkly 
contrasted with experimental data and independent QCD-based 
calculations~\cite{Garg:2013,Braun:1995,Cley:1997,Cley:2006,Kadam:2019,Gupta:2022}. 
Recently, the deconfinement transition 
of rotating hot and dense matter has been studied within the HRG model~\cite{Fujimoto:2021}. 
Here we shall explore the higher moments of fluctuations 
of net-baryon number, net-charge and net-strangeness for a rotating QCD medium and 
their dependence on the collision energy. We perform our analysis in the 
framework of a rotating HRG model to estimate the different thermodynamic 
quantities such as pressure, entropy density as well as susceptibilities and their ratios
~\cite{Fujimoto:2021,Ebihara:2017}. Calculations including the effects 
of rotation are shown for the ratios of quartic and quadratic (kurtosis), cubic and
quadratic (skewness) as well as quadratic charge fluctuations normalized to their 
mean value along a phenomenologically determined freeze-out curve in
HICs~\cite{Cley:2006,Karsch:2011}.

In Sec.~\ref{sec:s2} we describe a reformulation of the standard HRG model, as 
modified due to the inclusion of rotation. This is then used to compute the 
equation of state via thermodynamic variables like pressure, entropy density and
 energy density. We report the main results with regard to the susceptibilities and 
their ratios in Sec.~\ref{sec:s3}. Their phenomenological utility via connection
to the observable moments of the conserved charge distributions is discussed as well.
Finally in Sec.~\ref{sec:s4}, we summarize our findings.

\section{\label{sec:s2}Rotating HRG model: Bulk properties}

In the HRG model~\cite{Braun:1995}, the confined phase of QCD matter is
modeled by an ideal relativistic gas of all known hadrons and resonances.
It has been used alongside ab initio lattice QCD treatments and both are
found to be mutually consistent in their overlap region of validity.

Under the coordinate transformation suitable for a rigid global rotation, all
local quantities can be expressed as functions of the co-rotating coordinates,
$x^{\mu}$, in the non-inertial rotating frame of reference instead of $\tilde{x}^{\mu}$
in the rest (lab) frame. The corresponding metric can be read as  
\begin{equation}
 g_{\mu \nu} = \eta_{a b} \dfrac{\partial \tilde{x}^a }{ \partial x^\mu } \dfrac{\partial \tilde{x}^b }{ \partial x^\nu}=
\begin{pmatrix}
1-(x^2+y^2)\omega^2 & y\omega & -x\omega & 0\\
y\omega & -1 & 0 & 0\\
-x\omega & 0 & -1 & 0\\
0 & 0 & 0 & -1
\end{pmatrix},
\end{equation}
with the Minkowskian metric taken as $\eta = \text{diag}(1,-1,-1,-1)$. To deal
with the fermions we introduce the vierbein, $\eta_{a b} = e^{\mu}_{a}  e^{\nu}_{b} g^{\mu \nu}$, and adopt   
\begin{equation}
\begin{aligned}
e^{t}_{0}=e^{x}_{1}=e^{y}_{2}=e^{z}_{3}=1,~~~~ e^{x}_{0}=y \omega,~~~~ e^{y}_{0}=-x \omega,
\end{aligned}
\end{equation}
taking all other components zero.

The explicit calculations~\cite{Fujimoto:2021} yield the pressure $p_{i}^{B/M}$ 
for $i^{th}$ baryon (superscript $B$, upper signs) or meson 
(superscript $M$, lower signs), and is given by

\begin{eqnarray}
\begin{aligned}
 p_{i}^{B/M}=&\pm  \frac{T}{8\pi^{2}} \sum_{l=-\infty}^{ \infty }\int dk_{r}^{2} \int dk_{z}\\
 &\sum_{\nu=l}^{l+2s_{i}}  J_{\nu}^{2}(k_{r}r) \ln(1 \pm e^{-(E_{i}-\mu_{i})/T}) .
\end{aligned}
\label{HRG1}
\end{eqnarray}
Here $J_{\nu}^{2}(k_{r}r)$ is the (squared) Bessel function and 
$\mu_{i}=B_{i}\mu_B+S_{i}\mu_S+Q_{i}\mu_Q$ is the total chemical potential with 
$B_i$, $S_i$ and $Q_i$ the respective baryon  number, strangeness and electric-charge
of the $i^{th}$ particle and the $\mu_B$, $\mu_S$ and $\mu_Q$ are the corresponding
chemical potentials. In order to implement the conservation laws for strangeness and electric 
charge, we have introduced the strangeness and the electric charge chemical 
potentials for the complete system. $\mu_S$ and $\mu_Q$ have non-zero (finite) 
values to obtain net strangeness equal to zero and baryon number 
to charge ratio $N_B/N_Q \simeq$ 2.52, which comes from 
beta-equilibrium with the Coulomb interaction in the heavy-ion collisions. The 
single-particle energy levels are given as
\begin{equation}
E_{i}= \sqrt{k_{r}^{2}+k_{z}^{2}+m_{i}^{2}}-(l+s_{i})\omega ,
\label{HRG2}
\end{equation}
where $s_{i}$ and $m_i$  are the spin and mass of the $i^{th}$ particle respectively. 
The final term behaves as an effective chemical potential due to rotation. We have 
incorporated all hadrons listed in particle data book~\cite{PDG:2014} up to an 
ultraviolet mass cut of $\Lambda=1.5$ GeV to save numerical cost~\cite{Cley:2006}. 

The causality condition prohibits the formation of unphysical condensates by forbidding 
the $(l+s)\omega$ term to become greater than the `free-particle' part of the energy 
dispersion $\sqrt{k_{r}^{2}+k_{z}^{2}+m_{i}^{2}}$. Imposing a boundary condition on 
the wave-function to be normalized within $r \le R$, the causality requires $R\omega 
\le 1$. The leading discretization effect of momenta from causality bound is in the 
low momentum region and thus an infrared cut-off for the $k_r$ integration is 
introduced~\cite{Fujimoto:2021}, which is defined as, 
$\Lambda_l^{IR}=\zeta_{l,1}\omega$.
Here $\zeta_{l,1}$ is the first zero of the Bessel function:
$J_l(\zeta_{l,1})=0$.
The $k_r$ integration in Eq.~\ref{HRG1} is then redefined as,
\begin{equation}
\int dk_r^2 \rightarrow \int_{(\Lambda_l^{IR})^{2}} dk_r^2 .
\end{equation}
\begin{figure}
\includegraphics[width=0.5\textwidth]{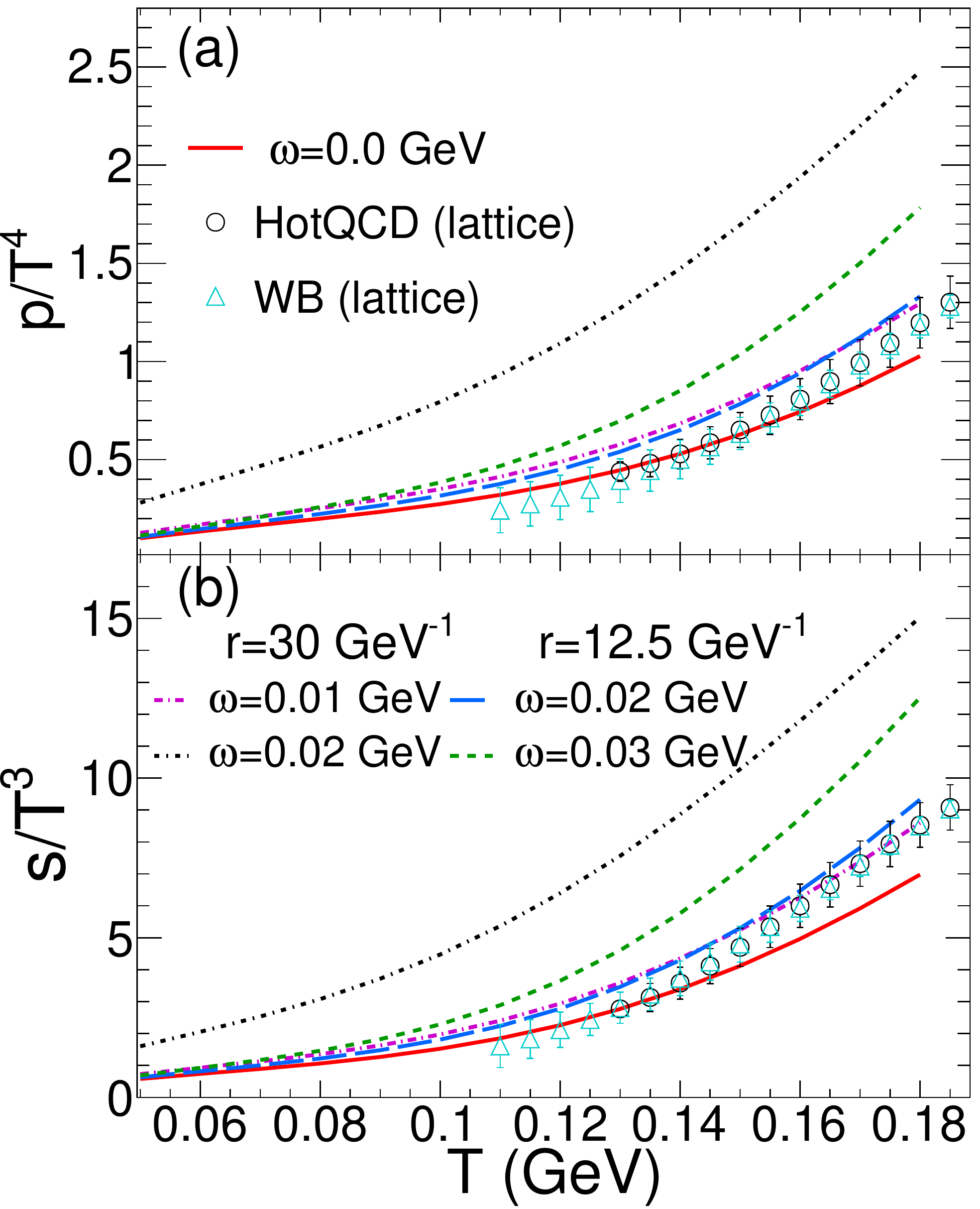}
\caption{\label{fig:pht1}(Top) Scaled pressure $p/T^4$ and (bottom) entropy
  density $s/T^3$ as a function of the temperature $T$ for different angular velocity 
$\omega$ at $\mu=0$. The open circle and cyan triangle symbols are for 
HotQCD~\cite{HotQCD:2014kol} and WB~\cite{Borsanyi:2013bia} lattice data 
respectively calculated at $\mu$ = 0. }
\end{figure}
Figure~\ref{fig:pht1} shows the effect of rotation ($\omega$) on 
thermodynamic variables like pressure $p/T^4$ (top panel) and entropy density 
$s/T^3$ (bottom panel) as a function of temperature $T$. The HRG results with 
and without rotation are compared with the lattice QCD results 
of the Hot-QCD Collaboration~\cite{HotQCD:2014kol} and the Wuppertal-Budapest (WB) 
Collaboration~\cite{Borsanyi:2013bia} shown in the same figure. Here we have 
considered two different centrality classes in HICs: 
(i) central collisions  where the fireball radius $R=30$ GeV$^{-1}$ (corresponds to 
$R=6$ fm) and (ii) peripheral collisions where $R=12.5$ GeV$^{-1}$ (corresponds to 
$R=2.5$ fm)  ~\cite{STAR:2017prc}. In Fig.~\ref{fig:pht1}, we have used
$r=30$ GeV$^{-1}$ ($r=12.5$ GeV$^{-1}$) for central (peripheral) collisions 
respectively. It is observed that in both cases, the pressure and entropy density 
increases with rotation $\omega$. The value of the angular velocity 
$\omega=0.01-0.04$ GeV, considered in this study, lies is in the range that is 
expected at freeze-out in  heavy-ion collisions~\cite{JiangLinLiao:2016prc}.
The causality condition $R\omega\le 1$ is respected too, for this range. 
It is observed that $p/T^4$ and $s/T^3$ increase with $T$ and there is a rising 
trend for both with increasing radius $r$. Similar behavior was also noted in 
Ref.~\cite{Fujimoto:2021}. The lattice results at lower $T < $ 150 MeV are 
well described by ideal HRG without taking rotation into account. 
The two graphs which correspond to $r=12.5$ GeV$^{-1}$ (blue 
dash) and $r=30$ GeV$^{-1}$ (black dotted) for $\omega=0.02$ GeV are shown in 
Fig~\ref{fig:pht1}. In this study, we consider the highest possible $r$ ($r\le R$) 
for both peripheral ($r=12.5$ GeV$^{-1}$) and central ($r=30$ GeV$^{-1}$) collisions 
to investigate the maximum effect of rotation $\omega$ on the fluctuations 
(susceptibilities).  
\begin{figure*}[ht!]
\includegraphics[width=1.0\textwidth]{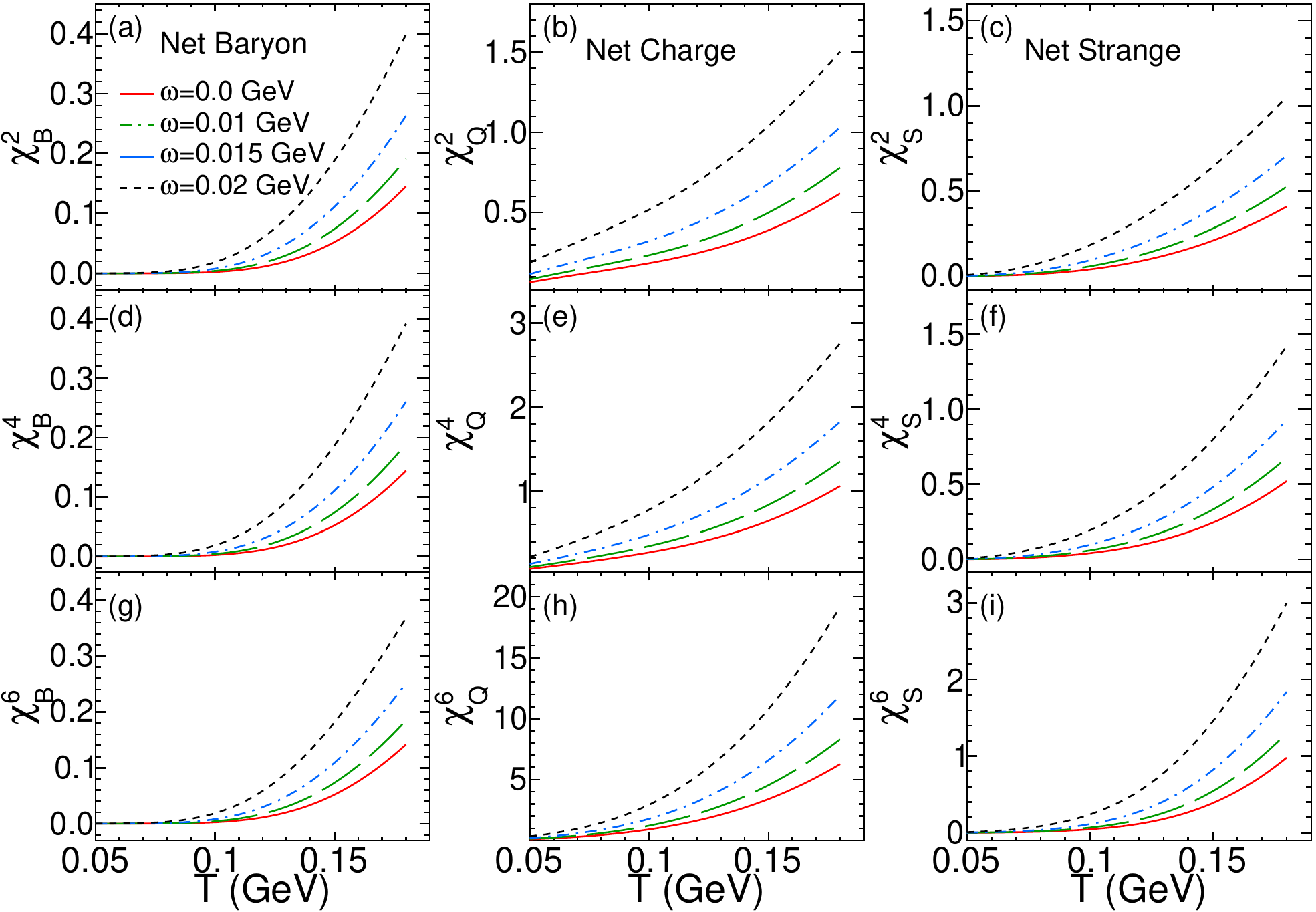}
\vspace{-0.2cm}
\caption{\label{fig:sustemp}$\chi^2$(upper row), $\chi^4$ (middle row) and 
  $\chi^6$ (lower row) as a function of the temperature $T$ for different
  angular velocity $\omega$ at $\mu=0$ for baryon number (left column),
  electric charge (middle column) and strangeness (right column).}
\end{figure*}

\section{\label{sec:s3}Observables: Moments and Susceptibilities}
In HRG model calculations, susceptibilities are 
defined as derivatives of 
the (scaled) pressure $p=-f$ with respect to the (scaled) chemical potential in the 
following way 
\cite{Bhatt:2016}
\begin{equation}
\begin{aligned}
\chi^{k}_{x}=\frac{\partial^{k}(\Sigma_{i}p_{i}/T^{4})}{\partial(\mu_{x}/T)^k},
\end{aligned}
\label{Eq:sus}
\end{equation}
where $k$ is the order of derivatives and the conserved quantum numbers such as 
baryon number, strangeness, and electric-charge are represented by $x$. 
From the definition of the grand canonical partition function, one can show that 
susceptibilities are related to the cumulants of the event-by-event multiplicity 
distributions which are measurable in the heavy-ion experiments via the relations,
\begin{eqnarray}
\chi^{1}&=&\frac{1}{VT^{3}}\langle 
N\rangle,~\chi^{2}=\frac{1}{VT^{3}}\langle(\Delta N)^2\rangle,
\chi^{3}=\frac{1}{VT^{3}}\langle(\Delta N)^3\rangle, \nonumber\\
 \chi^{4}&=&\frac{1}{VT^{3}}\langle(\Delta N)^4_c\rangle
\equiv(\langle(\Delta N)^4\rangle-3\langle(\Delta N)^2\rangle^2),\nonumber
\end{eqnarray}
\begin{eqnarray}
\chi^{5}&=&\frac{1}{VT^{3}}\langle(\Delta N)^5_c\rangle \nonumber \\
 &\equiv& \langle(\Delta N)^5\rangle-10 \langle (\Delta   
   N)^3\rangle\langle(\Delta N)^2\rangle, \nonumber\\ 
\chi^{6}&=&\frac{1}{VT^{3}}\langle(\Delta N)^6_c\rangle \nonumber \\   
&\equiv& \langle(\Delta N)^6\rangle-15 \langle (\Delta   
   N)^4\rangle\langle(\Delta N)^2\rangle-10\langle(\Delta N)^3\rangle^2 \nonumber\\
   &+&30\langle(\Delta N)^2\rangle^3~.
\label{Eq:chi} 
\end{eqnarray}
Here, $N$ is the number of the measured particles or conserved charges. $\Delta 
N=N-\langle N \rangle$ is the fluctuation around the event-averaged mean $\langle 
N\rangle$. The ratios of the susceptibilities yield the products of moments as,
\begin{figure*}[ht!]
  \includegraphics[width=1.0\textwidth]{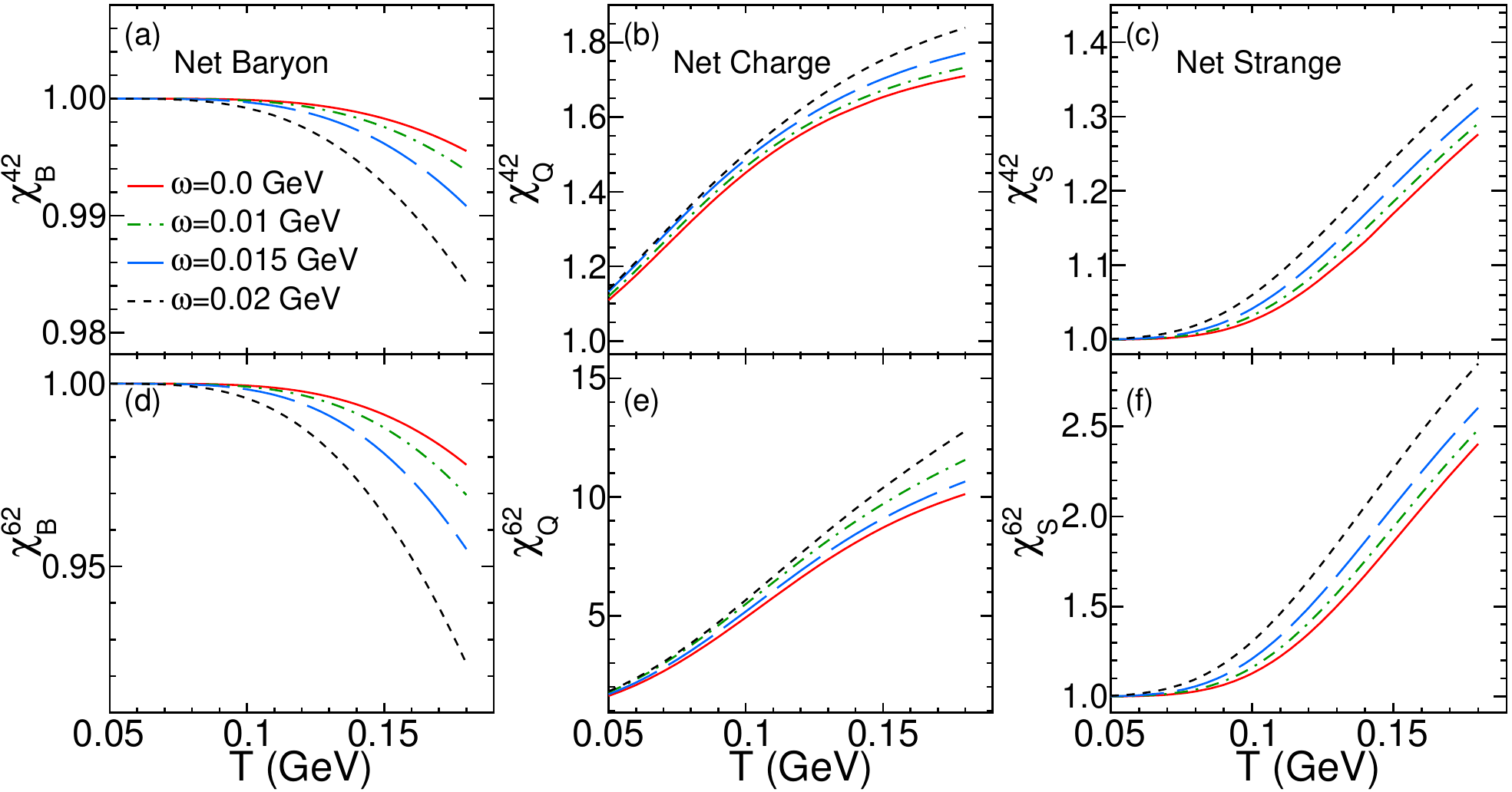}
  \caption{\label{fig:susrtemp} $\chi^{42} = \chi^4/\chi^2$ (top row) and 
  $\chi^{62} = \chi^6/\chi^2$ (bottom row) as a function of the temperature $T$ 
for different angular velocity $\omega$ at $\mu=0$ for baryon number (left column), 
electric charge (middle column) and strangeness (right column).}
\end{figure*}

\noindent
\begin{equation}
\frac{\chi^2}{\chi^1}=\frac{\sigma^2}{M},~~~\frac{\chi^{3}}{\chi^{2}}=S\sigma,~~~ 
\frac{\chi^{4}}{\chi^{2}}=\kappa \sigma^{2}, ~~~\frac{\chi^{6}}{\chi^{2}}=\kappa^H
\sigma^{4}.
\label{Eq:moments}
\end{equation}
The mean $M$, the variance $\sigma^2$, the skewness $S$, the kurtosis $\kappa$ and 
the hyper-kurtosis $\kappa^H$ are obtained experimentally from the measured 
event-by-event multiplicity distributions~\cite{STAR:2022vlo,Bazavov:2020bjn}. 
These moments characterize the shape of the multiplicity distributions.
Equation~\ref{Eq:moments} establishes the relation between the experimentally 
measurable moments and the theoretically calculable susceptibilities from 
thermodynamics and this relation is to be utilized to provide the predictions 
from our modified HRG model. One advantage of measuring the $\sigma^2/M$, $S\sigma$, 
and $\kappa\sigma^2$ is that the volume dependence of $M$, $\sigma$, $S$, and 
$\kappa$ cancel out in the ratios; hence theoretical calculations can be
directly compared with the experimental measurements. 

\section{\label{sec:s4}Results and Discussions}

The susceptibilities are related to ensemble fluctuations of the conserved
quantities which can be obtained from event-by-event multiplicity measurements
in heavy-ion collision experiments. The study of fluctuations in heavy-ion
  collisions is an important tool for the experimental determination of
  the QCD critical point and the first-order phase transition.
Figure~\ref{fig:sustemp} shows the variation of second-order ($\chi^2$), 
fourth-order ($\chi^4$) and sixth-order ($\chi^6$) susceptibilities of different 
conserved charges: baryon number (left column),  electric charge (middle column) and 
strangeness (right column) with temperature $T$ at chemical potential $\mu=0$ for 
different values of the angular velocity $\omega$. The susceptibilities are
  calculated along the freeze-out curve determined from the universal freeze-out 
condition on a fixed value of entropy density $s/T^3\simeq$ 7.0. The variation of even order
susceptibilities corresponding to the conserved charges are similar. However, the 
absolute values of the higher order susceptibility are generally larger compared to
the second order susceptibilities due to the larger weight for higher order. 
If the chiral crossover is near the freeze-out line on the phase diagram 
the higher order cumulants for net baryon number and electric charge 
fluctuations should show a sharp contrast with the HRG reference values~\cite{Friman:2011pf}.

The equation of state is clearly affected due to rotation (for example, the 
pressure is shown to behave in accordance with the expected centrifugation as a 
function of radial distance from the axis of rotation in Ref.~\cite{Fujimoto:2021}, 
also see Fig.~\ref{fig:pht1}). 
By virtue of this influence the susceptibilities also inherit the effects of 
rotation as observed in our results. The best way to see how rotation impacts these 
observables is to consider the whole system to be composed of an infinite sequence 
of coaxial shells or cylindrical surfaces with radii $r_{i} \le R, i=1,2,3,...,$ and 
regard one such cylindrical surface at fixed radial distance, $r$, as a member of 
the sub-ensemble (of the ensemble of identical cylindrical systems of radius $R$). 
Each cylindrical shell is in thermal and chemical (grand canonical) contact and 
equilibrium with the solid inner core and the remaining outer annular parts of the 
cylindrical system simultaneously. It is expected that the centrifugation from 
rotation and boundary effects due to the causality bound will lead to the exchange of 
conserved charges between the inner core and annular cladding between which the thin 
shell at $r_i$ is sandwiched in a biased way or asymmetrically. This is the likely 
physical origin of the effects observed here and thus may be attributed to the 
geometry of the system and the kinematics of its rotation.
\begin{figure*}[hbt!]
  \includegraphics[width=1.0\textwidth]{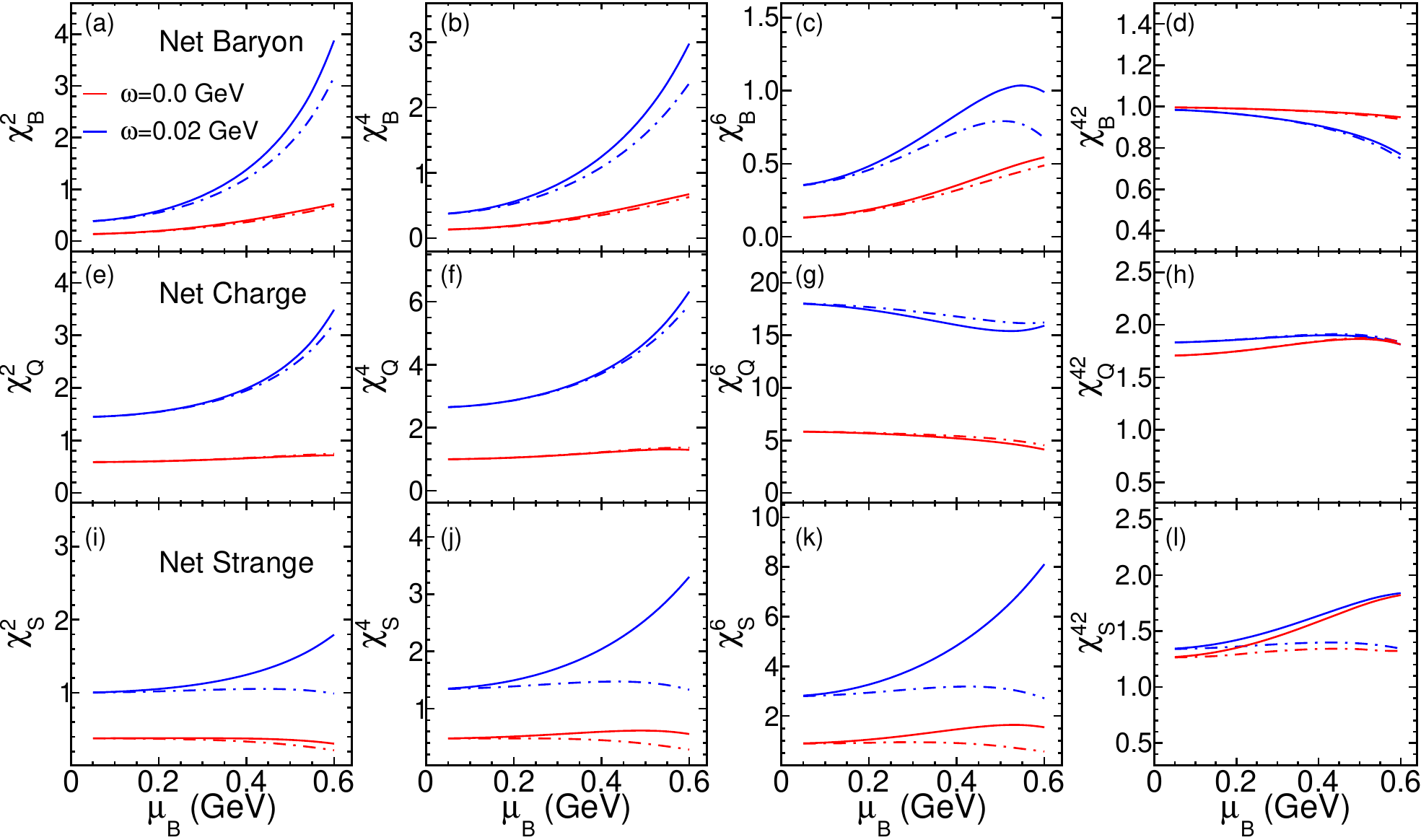}
\vspace{-0.2cm}
\caption{\label{fig:fig4}Susceptibilities ($\chi^2$, $\chi^4$, and $\chi^6$) and the 
ratios ($\chi^{42} = \chi^4/\chi^2$) for net-baryon, net-charge and net-strangeness 
as a function of $\mu_B$ without (solid line) and with (dashed line) inclusion of 
rotation.}
\end{figure*}
\begin{figure*}[hbt!]
   \includegraphics[width=1.0\textwidth]{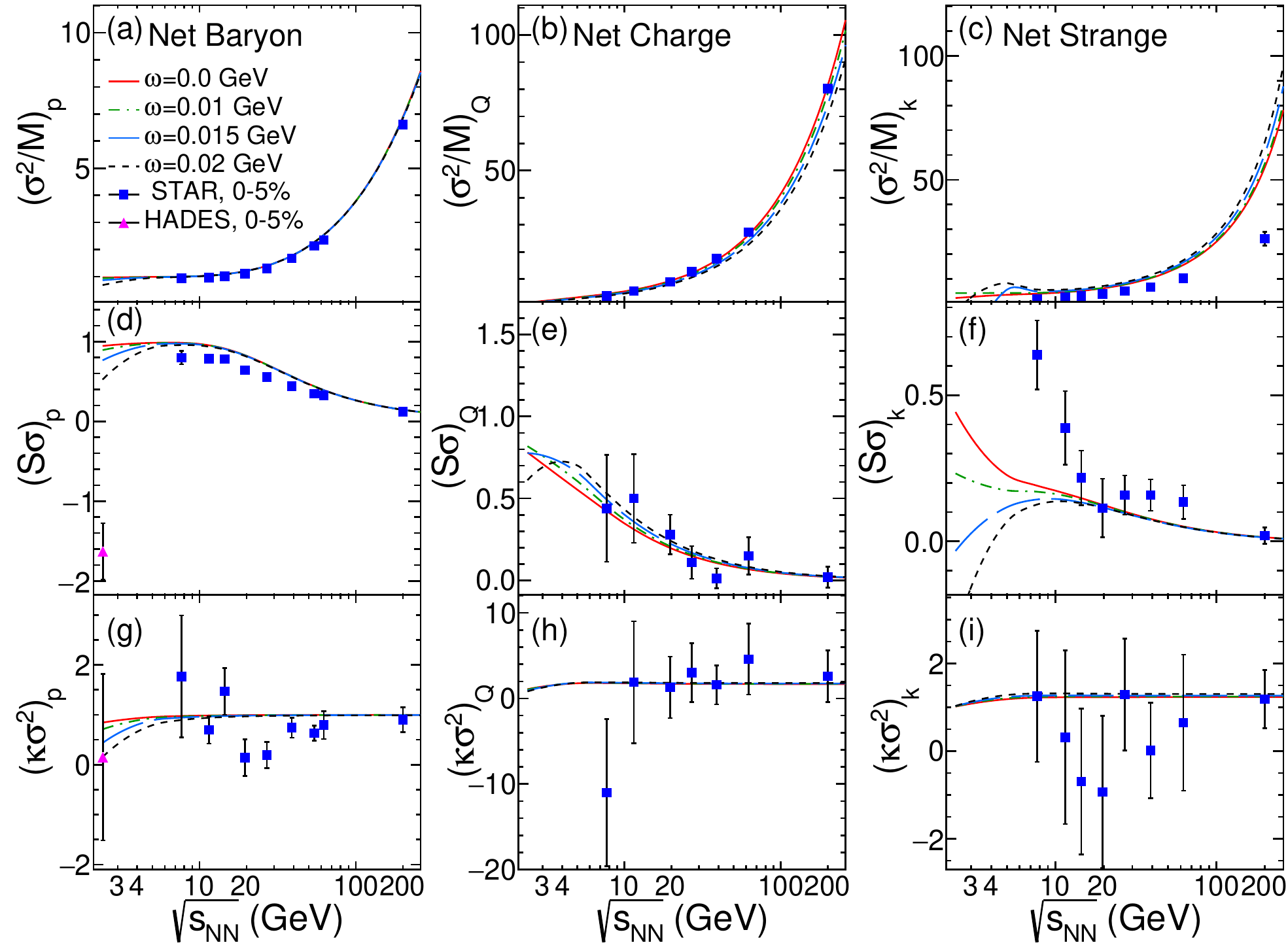}
\vspace{-0.2cm}
\caption{\label{fig:momcen}Products of moments of net-proton (left column),
  net-charge (middle column) and net-kaon (right column) as functions of the
  centre-of-mass energy for different angular velocity $\omega$.}
\end{figure*}

In case of HRG model with and without rotation, the susceptibilities increase with 
temperature $T$. At low temperature the contribution to $\chi_B$ is mainly from
proton and neutron. With increase of temperature $T$ other baryons also start 
contributing, hence with temperature $T$ susceptibilities $\chi_B^2$, 
$\chi_B^4$, $\chi_B^6$ increase. We also notice that, the susceptibilities  increase 
with rotation $\omega$ at specific temperature $T$. This is because, number density, 
pressure as well as entropy density increase with $\omega$ after a specific 
temperature $T$. Similar behavior with temperature $T$ and rotation $\omega$ is also 
observed for $\chi_Q$ and $\chi_S$.

\begin{figure*}[hbt!]
  \includegraphics[width=1.0\textwidth]{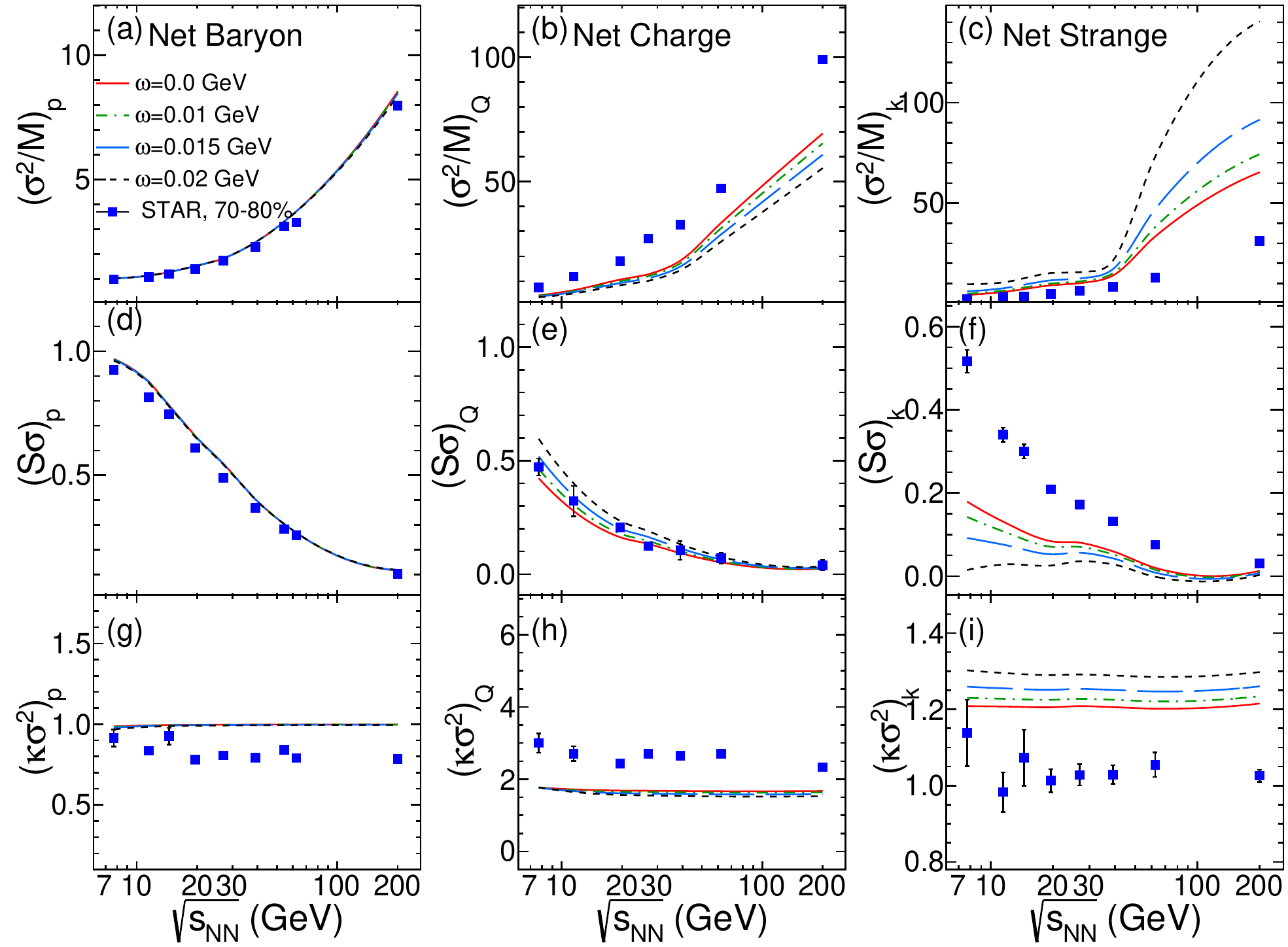}
  \caption{\label{fig:momper}Products of moments of net-proton (left column),
    net-charge (middle column) and net-kaon (right column) as functions of
    the centre-of-mass energy for different angular
  velocity $\omega$ for peripheral collisions.}
\end{figure*}
In case of $\chi_Q$, the major contributions are from pions and kaons. With increase 
in temperature more charged particles ($\rho$, $K^*, p, \Delta$, etc.) are produced in 
the system which contribute to the increase in susceptibilities at higher 
temperature. Similarly the dominant contribution to the ($\chi_S$) at low 
temperatures comes from 
kaons which have strange quantum number $\pm$1. As a result, magnitudes of the 
susceptibilities ($\chi^2_S$, $\chi^4_S$, and $\chi^6_S$) are similar in the low-$T$ 
region up to 0.1 GeV. At higher temperature, other strange hadrons (e.g., $\Sigma$, 
$\Xi$, $\Omega$, etc.) having strangeness number $\pm$2 and $\pm$3 contribute to the 
higher values of susceptibilities. Hence, at high-$T$, the higher order 
susceptibilities ($\chi^6_S$) increase rapidly. So, the rotation effect 
enhances the fluctuations of the conserved quantities along the freeze-out curve.
Figure~\ref{fig:susrtemp} shows the ratios of susceptibilities 
($\chi^4/\chi^2$, $\chi^6/\chi^2$) as a function $T$ for net-baryon, net-charge and 
net-strangeness for different values of $\omega$. The net-baryon 
susceptibilities ratios $\chi_B^4/\chi_B^2$, $\chi_B^6/\chi_B^2$ decrease as 
a function of $T$. The values are further lowered as we increase $\omega$. On 
the other hand, as in case of net-charge and net-strangeness shows the increasing 
trend as a function of $T$. The ratios are systematically higher with increase in 
$\omega$.

The charge conservation and zero strangeness condition plays an important 
role in the calculation of the fluctuations in presence of rotation particularly at 
lower collision energies (higher $\mu_B$)~\cite{Fujimoto:2021}. The 
strangeness chemical potential $\mu_S$ grows as $\mu_B$/3 and compare with the 
free-out data from the experimental measurements, it is crucial to impose the zero 
strangeness ($N_s$ = 0) and $N_B/N_Q \simeq$ 2.52 condition.
Figure~\ref{fig:fig4} shows the 
susceptibilities ($\chi^2$, $\chi^4$, and $\chi^6$) as a function of $\mu_B$ with 
and without inclusion of conservation effect along the freeze-out curve which is 
determined by imposing the fixed value of entropy density $s/T^3\simeq$ 7.0. 
For rotation $\omega = 0$, $\chi^2_B$, $\chi^4_B$, and $\chi^6_B$ are almost same as 
with and without charge conservation. At $\omega = 0.02$ GeV, the charge conservation 
decrease the baryon fluctuations as compared to without conservation. In case of 
$\chi_Q$, conservation has no effect on the charge fluctuations both with and 
without inclusion of rotation case. In case $\chi_S$, the fluctuations are large 
without charge conservation as compared to the with charge conservation and it is 
true for both zero and non-zero values for rotation. The charge conservation 
diminishes the strangeness fluctuations along the freeze-out curve. However, the 
charge conservation effect is not observed in the ratios of susceptibility 
($\chi^{42}$) except strangeness fluctuation case. 

Products of moments of net-proton, net-charge and net-kaon as functions of the 
centre-of-mass energy $\sqrt {s_{NN}}$ for central heavy-ion collisions $(0-5\%)$
at different angular velocity $\omega$ are shown in the Fig.~\ref{fig:momcen}. In 
order to make connections with experiments, the beam energy dependence of 
the temperature $T$ and the chemical 
potentials $\mu$'s has to be provided. We used the parametrization of temperature 
$T$ and chemical potential $\mu$'s with centre-of-mass energy $\sqrt {s_{NN}}$
along the freeze-out curve following Ref.~\cite{Cley:2006,Karsch:2011}. The values 
of $\omega$ considered here are $0.01,~0.015$ and $0.02$ GeV. 
Here, the net-proton fluctuations acts as proxy for net-baryon fluctuations 
and net-kaon acts as proxy for net-strangeness fluctuations.
The theoretical estimation of ratios of 
susceptibilities for different conserved charges are compared with the product of 
moments for net-proton, net-charge and net-kaon as a function of $\sqrt {s_{NN}}$
from measured data of STAR~\cite{STAR:2014prl,STAR:2021prla,STAR:2017tfy} and 
HADES~\cite{HADES:2020prc} experiment. We found that $(\sigma^{2}/M)_p$ has 
insignificant dependence on the  rotation $\omega$. For $(S\sigma)_p$ and 
$(\kappa\sigma^2)_p$, the effect of rotation is significant at lower $\sqrt {s}$ and 
the ratios decreases with increase of $\omega$ values. At the 
higher collision energies, the rotation has no effect on the ratios of net-proton 
susceptibilities. It should be noted  that, $(\kappa\sigma^2)_p$ with finite 
rotation for $\omega=0.02$ GeV, can explain the data of HADES while normal HRG fails 
to explain. The $(\sigma^2/M)_Q$ increases with respect to the center of mass 
energy for with or without inclusion of rotation.
The $(\sigma^2/M)_Q$ and $(S\sigma)_Q$ are weakly dependent on 
$\omega$ whereas $(\kappa\sigma^2)_Q$ is insignificant to rotation. In case 
of $(\sigma^2/M)_k$, rotation has small effect at higher $\sqrt {s_{NN}}$. For 
$(S\sigma)_k$ there is significant and non-monotonic effect of rotation observed 
 at lower collision energies. $(S\sigma)_k$ values decreases with the increase in 
rotation. $(\kappa\sigma^2)_k$ increases with the increase in rotation.

We have also estimated the ratio of susceptibilities for peripheral collisions.
From the causality condition $R\omega\le 1$, the values of angular velocity $\omega$ 
considered are $\omega=0.02,~0.03,~0.04$ GeV for $r<R=12.5$ GeV$^{-1}$. Ratio of 
susceptibilities from theoretical estimation are compared with product of moments 
from experimental measurements from STAR~\cite{STAR:2021prlb}. 
As the freeze-out parametrization used above, was suitable for  central 
collisions, we have used the parameterized values of $T$, $\mu$'s at different $\sqrt 
{s_{NN}}$ for peripheral collisions $(70-80\%)$ from Ref.~\cite{STAR:2017prc}. 
Figure~\ref{fig:momper} shows the comparison of theoretical estimation with 
experimental measurements of the product of moments for the peripheral collisions. 
 We don't observe the effect of rotation for ratios of net-proton number 
fluctuations for peripheral collisions. The behavior of ratios of the fluctuations 
for net-electric charge and net-strangeness are in peripheral collisions is 
completely different from what has been observed for central collisions. 
$(\sigma^2/M)_Q$ values increase with $\sqrt {s_{NN}}$, but decrease with the 
increase in rotation, particularly at higher collision energies. In case of 
$(S\sigma)_Q$, at lower collision energies the $S\sigma$ values increase with 
increase in rotation and at higher $\sqrt {s_{NN}}$ the effect of rotation is 
diminished. $(\kappa\sigma^2)_Q$ values are independent of rotation. The 
most pronounced effect is observed in case of net-kaon fluctuations. 
$(\sigma^2/M)_k$ values increase with the increase in rotation and the effect is 
more at higher energies, where as $(S\sigma)_k$ values decrease with the increase in 
rotation values of the system. $(\kappa\sigma^2)_k$ values increase with the 
increase in rotation and remain constant for all $\sqrt {s_{NN}}$.
To summarize the findings in peripheral collisions, it is observed that, the effect 
of rotation is significant for electric-charge and strangeness but much less so for 
ratio of susceptibilities for baryon number for the $\omega$ values considered 
($\omega = 0.02,~0.03,~0.04$ GeV).

We here briefly outline some future directions and potential extension of this work.
The ideal HRG model takes into account the attractive interactions 
between its hadronic constituents via inclusion of resonances and we have
analyzed the same idealized system rotating rigidly inside an infinite cylinder.
Further refinements to simulate the HIC fireball more realistically, such as including
finite size effects in the longitudinal direction, repulsive interactions, inhomogeneity~\cite{Chernodub:2020qah}, 
anisotropy, parallel magnetic field~\cite{Mukherjee:2023qvq},
repulsive interactions~\cite{Pradhan:2023rvf}, are possible and can be
built upon this work subsequently.
In the literature, the possibility of a sequential, flavor-dependent freeze-out 
has been explored~\cite{Ratti:2018ksb}. Considering such a scenario 
may help in a more refined interpretation of the reported results here 
leading to a better understanding of the freeze-out process in HICs.

The QCD phase transition in the presence of magnetic field has been 
studied extensively in the literature. Given the analogous properties
 of a magnetic field and rotation in the non-relativistic domain and 
certain similarities in relativistic cases~\cite{Mameda:2015ria} it is 
interesting, both theoretically and phenomenologically, to consider the possible consequences of having both acting
 simultaneously as should be the situation in off-central HICs. 

\section{\label{sec:s5}Summary and Conclusion}
The fluctuations of conserved numbers, namely net-baryon number, net-strangeness and 
net-charge were studied using HRG model in the presence of global rotation. We have 
chosen three different values of the angular velocity $\omega$ for both central and 
peripheral collisions in the expected range from simulation and measurement at 
RHIC~\cite{JiangLinLiao:2016prc, STAR:2017}. This study gives the  maximum limit of 
effect of rotation on the product of moments or fluctuations of conserved charges.  

The susceptibilities of baryon number, electric charge and strangeness number are 
found to be sensitive to rotation. We have studied the behavior of moments as a 
function of $\sqrt {s_{NN}}$ using  freeze-out parametrization for central 
collisions for $\omega=0$ as well as using parameters estimated for peripheral 
collisions. $(S\sigma)_B$ and $(\kappa\sigma^2)_B$ show a decreasing trend with 
increasing $\omega$ at very low centre-of-mass energy which could explain the data of 
$(\kappa\sigma^2)_p$ measured with  HADES at $\sqrt {s_{NN}}=2.4$ GeV.
We found that rotation, parameterized by $\omega$, has potentially important 
implications for the interpretation of heavy-ion collision data. In future studies, 
$\omega$  may  play the role of one of the freeze-out parameters besides $T$ and 
$\mu_B$.


\bibliographystyle{elsarticle-num}
\nocite{*}
\bibliography{fluctuation_epjc}

\end{document}